\documentclass[a4paper]{jpconf}
\usepackage{graphicx}

\usepackage{wrapfig}
\usepackage[abs]{overpic}

\usepackage{caption}
\usepackage[export]{adjustbox}
\usepackage{bm}
\usepackage{cite}

\newcommand{\twofig}{0.49\columnwidth}

\newcommand{\dptt}{$\delta p_\mathrm{TT}$}
\newcommand{\signal}{CC1p1$\pi^+$}

\begin{document}
\title{Measuring neutrino-induced exclusive charge-current final states on hydrogen at T2K}

\author{D. Coplowe, X.-G. Lu and G. Barr for the T2K Collaboration}

\address{Department of Physics, University of Oxford, Oxford, United Kingdom}

\ead{david.coplowe@physics.ox.ac.uk}

\begin{abstract}\\
By taking advantage of symmetries with respect to the plane containing the directions of the neutrino and outgoing lepton, it is possible to isolate neutrino interactions on hydrogen in composite nuclear targets. This technique enables us to study the `primary' neutrino-nucleon interaction and therefore gain access to fundamental model parameters free from nuclear effects. Using T2K Monte Carlo equivalent to $\sim7\times10^{21}$ POT, we present an update on the measurement of the exclusive charged-current $\mu^-$, p, $\pi^+$ final state on hydrogen.
\end{abstract}
\vspace{-1cm}

\section{Introduction}
If future long baseline neutrino experiments are to achieve systematic uncertainties of $\mathcal{O}(3\%)$~\cite{DUNE}, our knowledge of the nucleus in neutrino interactions needs to be improved. Uncertainties in the dynamics of the nucleus propagate through to oscillation analysis in two forms: neutrino energy reconstruction and in the evaluation of selection efficiencies for the respective signal and background~\cite{PRD91.072010.2515}. Isolating neutrino interactions on hydrogen~\cite{PRD.92.5.051302.2015, arXiv:1512.09042, arXiv:1606.04403} alleviates experimenters from many of these uncertainties, resulting in improved neutrino energy reconstruction and access to precise measurements of neutrino-nucleon interaction parameters such as the resonance axial mass.
\begin{wrapfigure}{r}{0.5\columnwidth}
\vspace{-0.4cm}
	\begin{overpic}[clip, trim = 45mm 0mm 0mm 0mm,width=.8\linewidth]{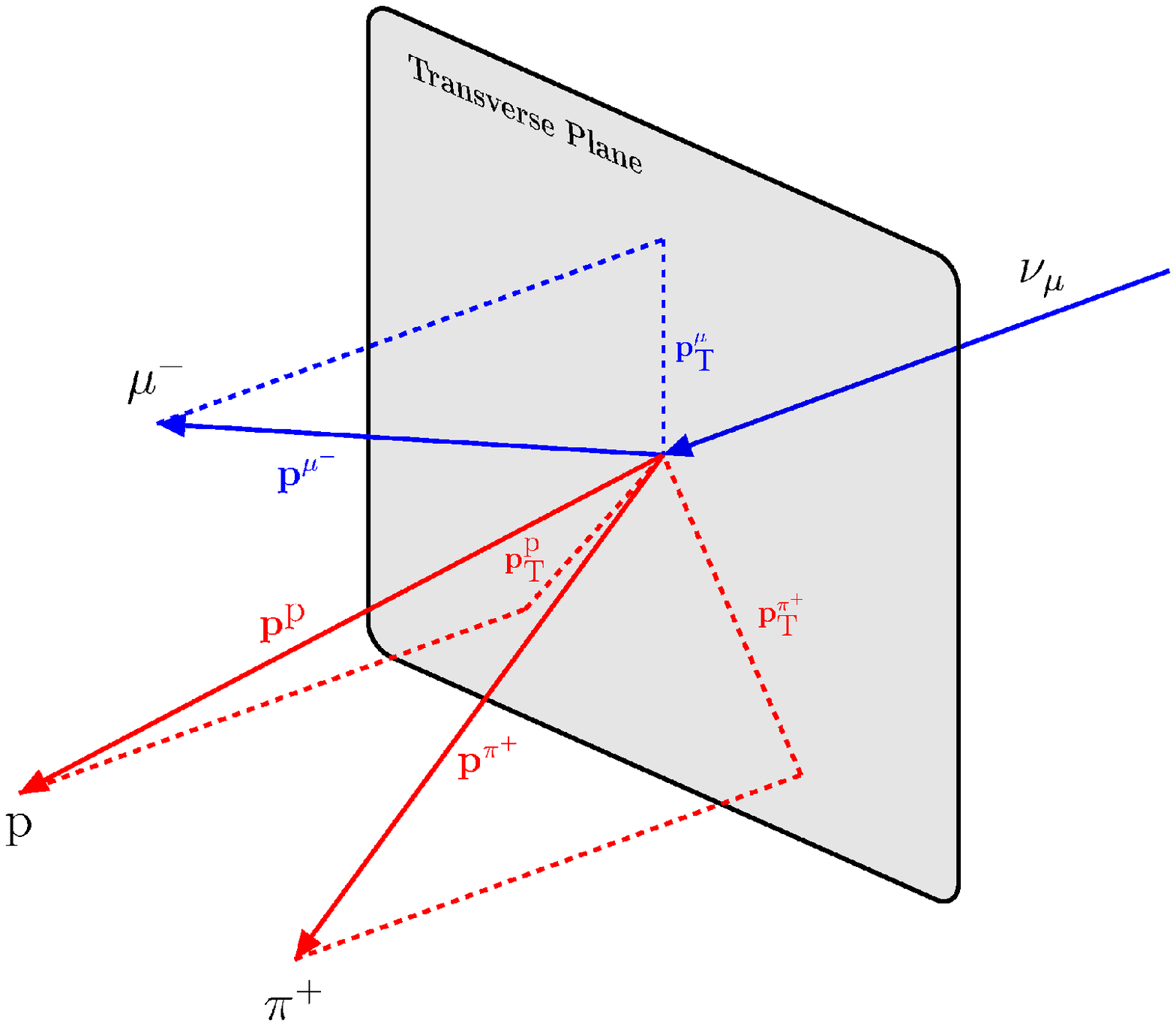}
		\put(175,-10){\includegraphics[width=0.26\linewidth]{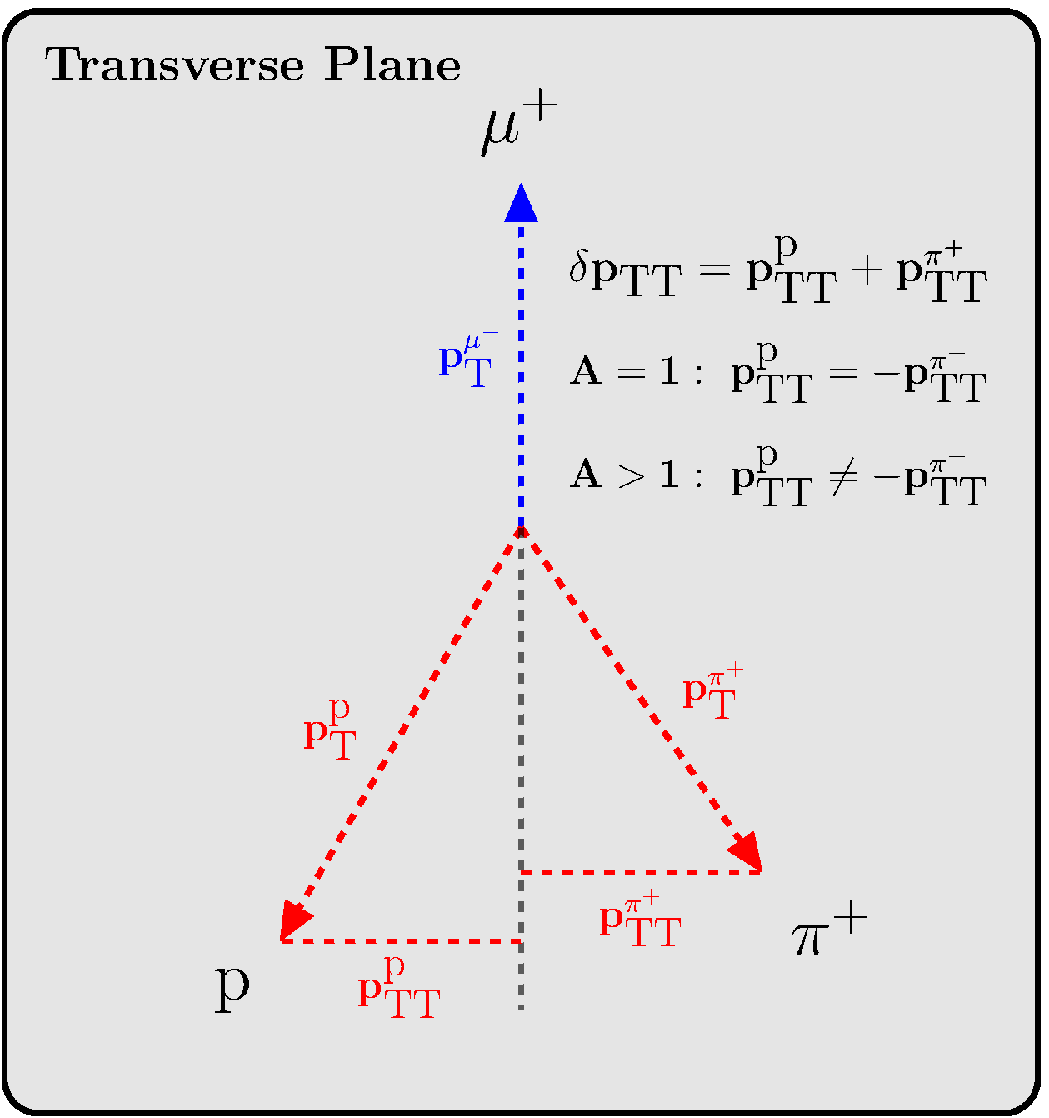}}
	\end{overpic}
\vspace{.3cm}
\caption{Double transverse momentum.}
\label{fig:dpTT_schem}
\vspace{-.4cm}
\end{wrapfigure}
\indent Following on from~\cite{1605.00154v1} which outlined the selection and reconstruction of exclusive charged current final states containing a single muon, proton and $\pi^+$ (\signal), and discussed the implementation of the double transverse momentum variable, \dptt~\cite{PRD.92.5.051302.2015}, used to isolate interactions on hydrogen at T2K's near detector, ND280, we present an update on the analysis. In~\cite{1605.00154v1}, we outlined a means of extracting the hydrogen signal by fitting functional forms to the signal and background, here, we discuss an alternative approach outlined in~\cite{PRD.93.112012.2016} for determining the signal component.

As seen in Fig.~\ref{fig:dpTT_schem}, in the absence of nuclear effects in which interactions take place on a free nucleon, $A=1$, \dptt~is balanced whereas bound nucleons, $A>1$,  have some initial isotropic Fermi motion and can undergo final state interactions  resulting in \dptt~being imbalanced. A full discussion of \dptt~can be found in~\cite{1605.00154v1, PRD.92.5.051302.2015, arXiv:1606.04403,arXiv:1512.09042}. Given the variables ability to extract events on hydrogen, our focus here is on the signal extraction for an absolute cross section where flux, detector and model systematics are included. Finally, the \signal-hydrogen energy dependence is considered.
\vspace{-0.2cm}
\section{Signal Extraction}
\begin{figure}[tb]
\centering
\includegraphics[width=\twofig]{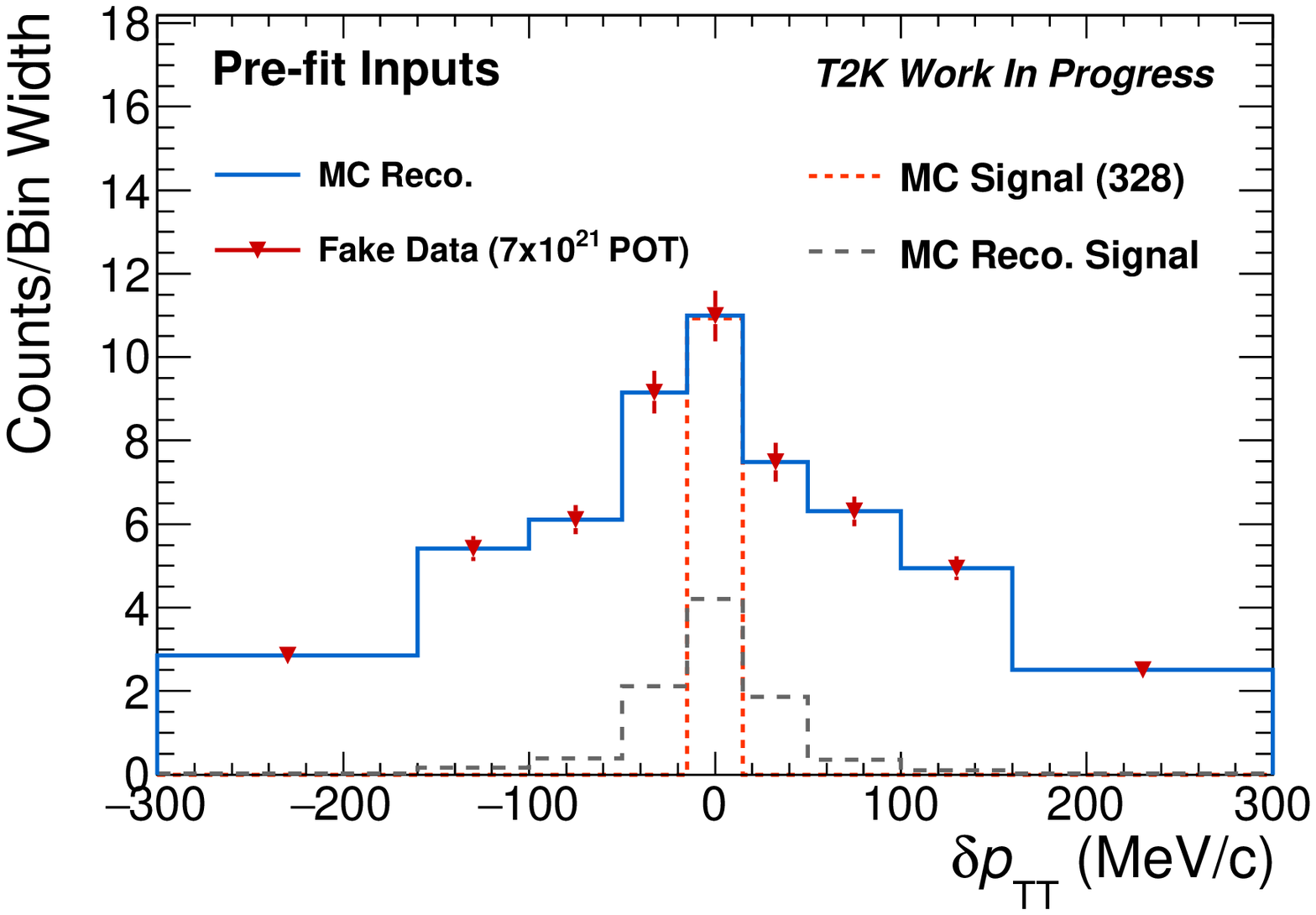}
\includegraphics[width=\twofig]{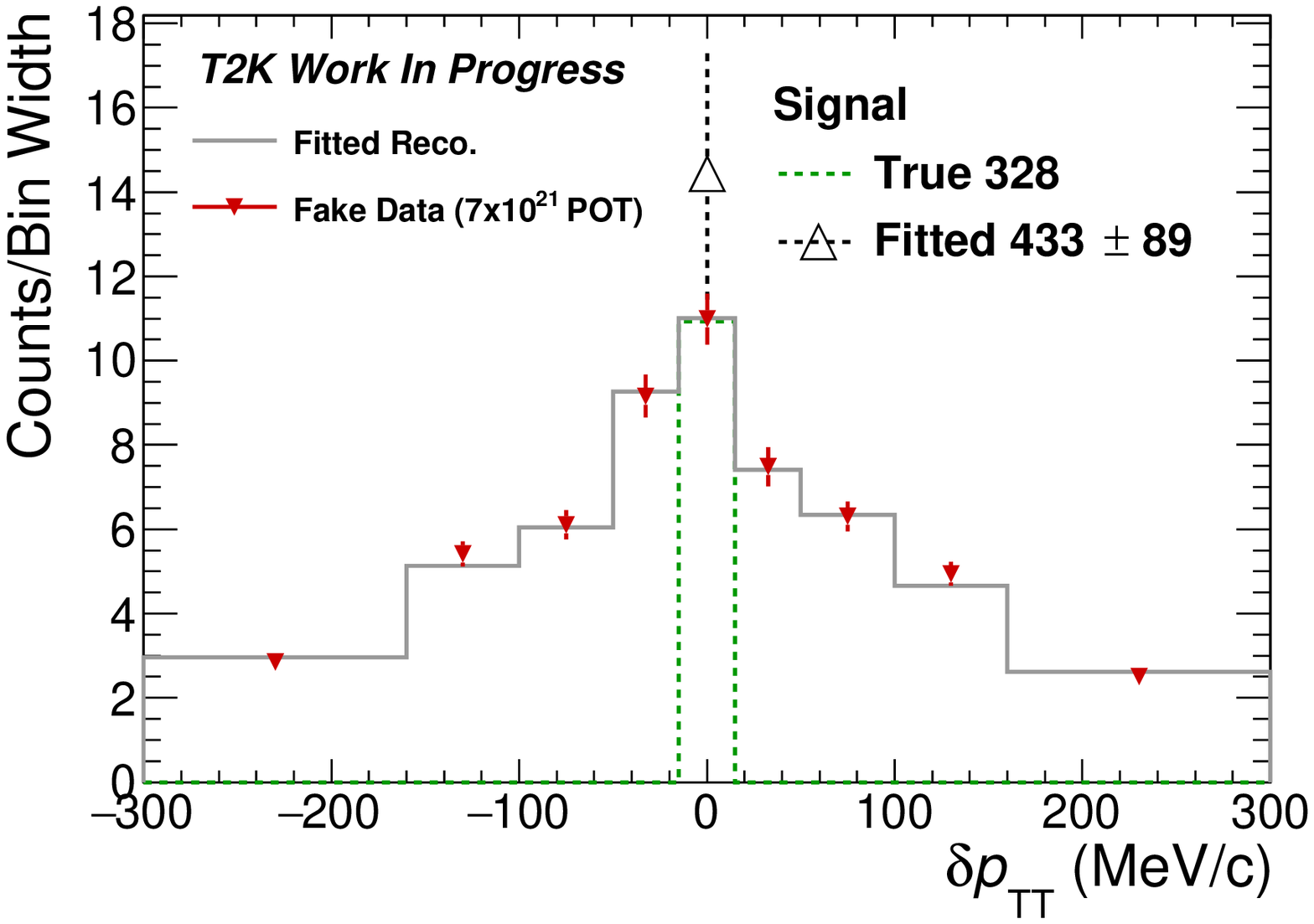}
\caption{Absolute cross section pre-fit inputs used to extract hydrogen component (\emph{left}) and fitted output compared to actual signal (\emph{right}).}
\label{fig:absxs}
\vspace{-0.8cm}
\end{figure}
The signal extraction technique is based on the likelihood method~\cite{PRD.93.112012.2016} that enables both model and detector systematic uncertainties to be included in the fit as nuisance parameters. In this approach we are fitting the number of selected events as a function of \dptt~(and $E_\nu$ for energy dependent case) by minimising
\vspace{-0.2cm}
\begin{equation}
\chi^2 = \chi^2_\mathrm{stat} + \chi^2_\mathrm{syst} = \sum^\mathrm{reco~bins}_j 2\left(N^\mathrm{exp}_j - N^\mathrm{obs}_j + N^\mathrm{obs}_j\ln\frac{N^\mathrm{obs}_j}{N^\mathrm{exp}_j}\right) + \chi^2_\mathrm{syst}
\vspace{-0.2cm}
\end{equation}
where $N^\mathrm{obs}_j$ is the number of observed and $N^\mathrm{exp}_j$ is the number of expected events in reconstructed bin $j$ and $\chi^2_\mathrm{syst}$ is the penalty term for the model and detector systematics. In the fit, the parameters of interest, $c_i$, which are encompassed in the $N^\mathrm{exp}_j$ term are adjusted to best fit the MC hydrogen-\signal cross section $(N^\mathrm{MC~signal}_i)$ given the data. The $N^\mathrm{exp}_j$ term reads
\vspace{-0.2cm}
\begin{equation}
N^\mathrm{exp}_j = \sum^{\mathrm{true~bins}}_i\left\{c_iN^\mathrm{MC~signal}_i + \sum^{bkg~reactions}_k  N^{\mathrm{MC~bkg}~k}_i\prod^\mathrm{model}_\alpha \omega(a_\alpha)^k_{ij}\right\}t^\mathrm{det}_{ij}r^\mathrm{det}_j\sum^{E_\nu}_n \omega^i_nf_n.
\vspace{-0.2cm}
\label{eqn:Nexp}
\end{equation}
Here $i$ runs over `true' \dptt~(and $E_\nu$ for energy dependent case) bins prior to detector effects, $k$ considers all background reactions. Theoretical uncertainties in the background predictions are included by the product of weighting functions, $\omega(a_\alpha)^k_{ij}$ for a given $k$ background. These functions account for variations in `true' and reconstructed \dptt~as a result in our knowledge of the theoretical model parameter $a_\alpha$. Detector effects are described by the matrix, $t^\mathrm{det.}_{ij}$, which maps true (i) to reconstructed (j) bins. The nuisance parameter vectors for model ($a_\alpha$) and detector ($r^\mathrm{det.}_j$) systematics are constrained by prior covariance matrices.
The final term, $\sum^{E_\nu}_n \omega^i_nf_n$ encompass flux dependences associated to each true bin. 
\vspace{-0.2cm}
\section{Results}
Presented is the outcome of MC data to MC data fits using the likelihood method defined above. In Fig.~\ref{fig:absxs}, the inputs prior to fitting (\emph{left}) and the fitted output (\emph{right}) for the absolute cross section are given. Here the `true' MC signal is shown along with its reconstructed counterpart and the full reconstructed MC. We include uncertainties from detector, model and flux systematics. Unique to this technique of isolating nuclear effects is that there is only one signal bin in each \dptt~distribution as seen in the MC signal, leaving only one $c_i$ to be determined. It is clear from the fitted result we are currently overestimating the signal, however this bias is expected to be removed once the implementation of model uncertainties are re-evaluated. One should note that the dominant uncertainties are coming from the detector systematics but these have not yet been optimised for this absolute cross section measurement. 

\begin{figure}[tb]
\begin{center}
\begin{minipage}{0.45\textwidth}
\begin{center}
\includegraphics[clip, trim = 0mm 3mm 0mm 15mm, width=0.9\textwidth]{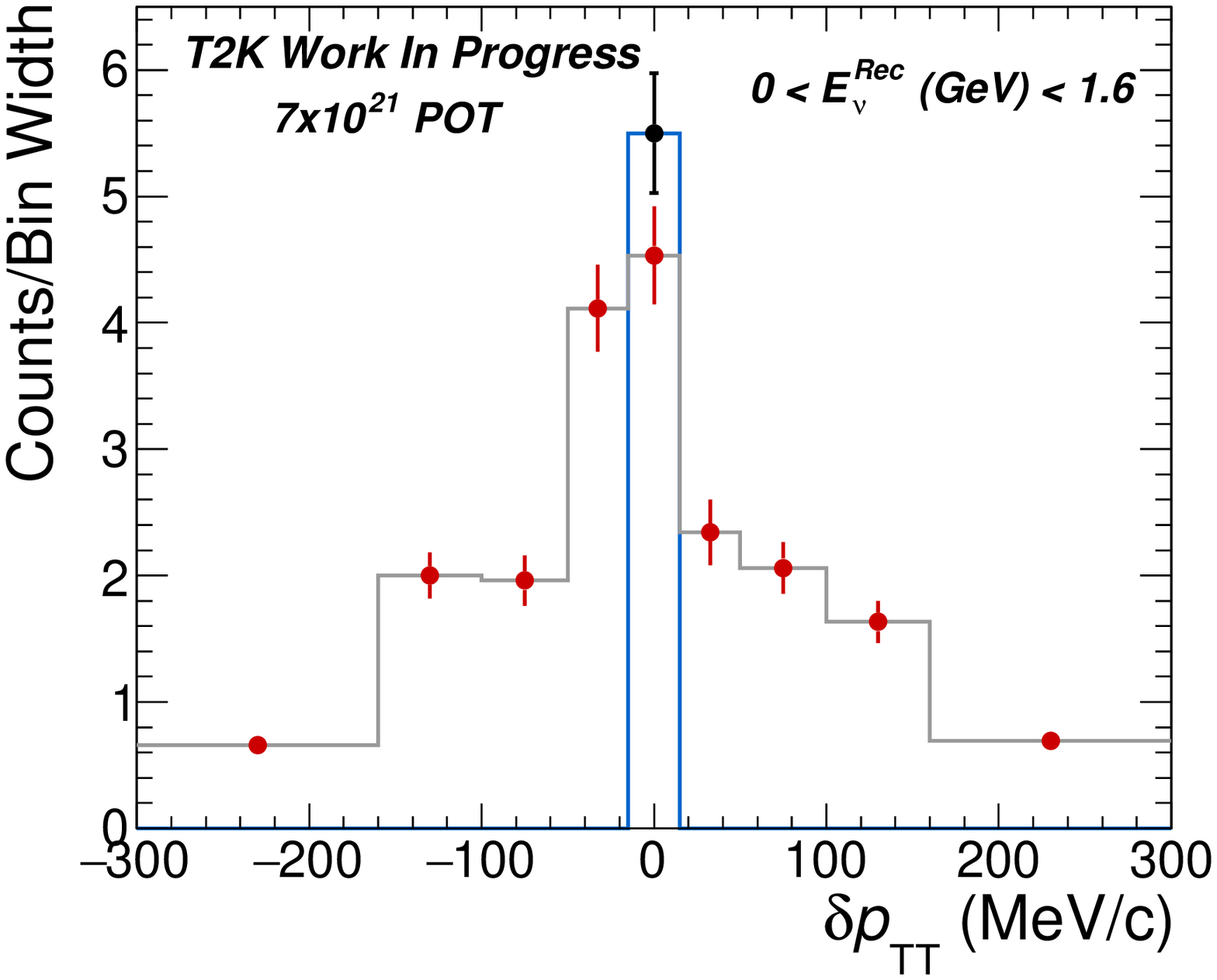}
\includegraphics[clip, trim = 0mm 3mm 0mm 15mm, width=0.9\textwidth]{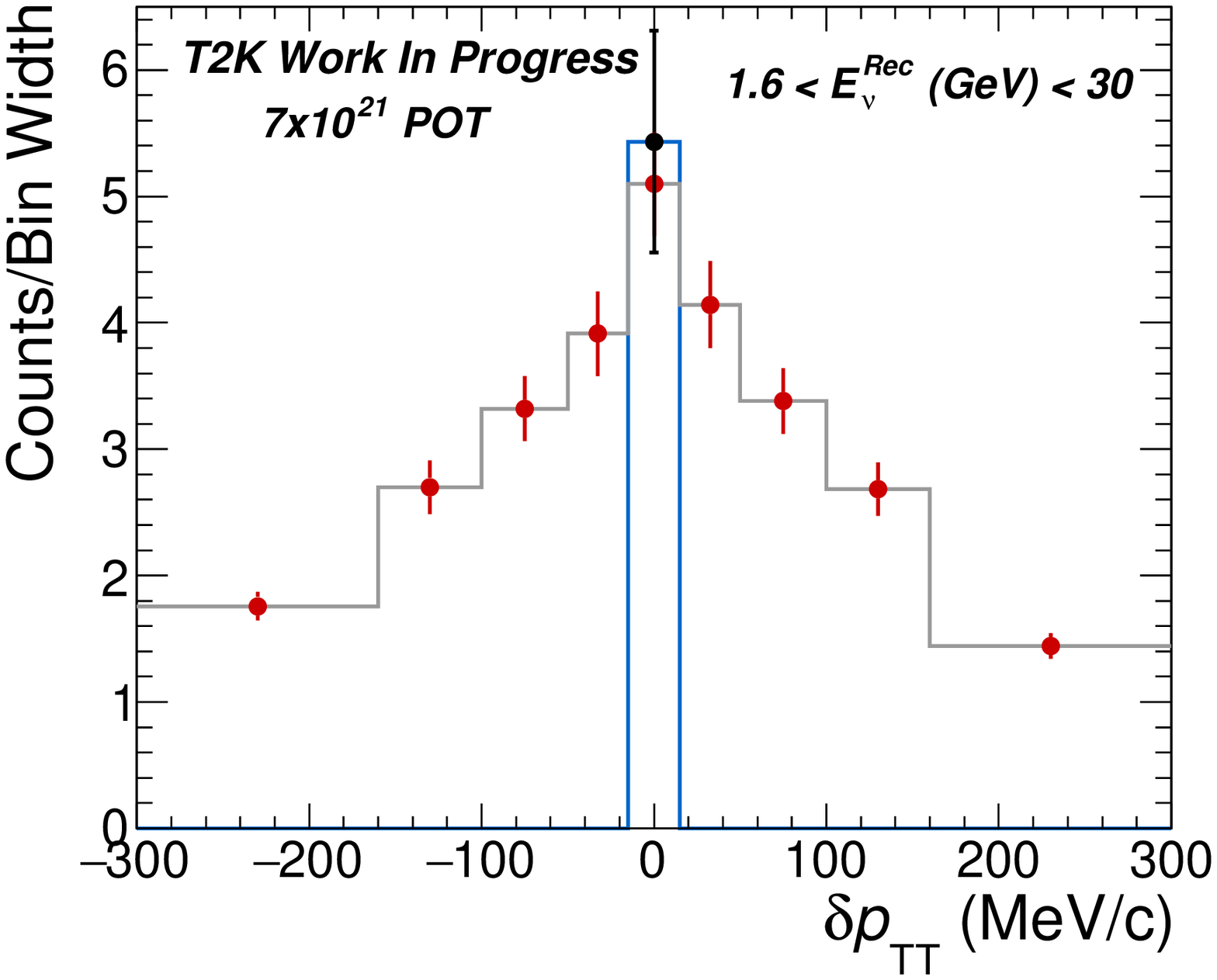}
\end{center}
\end{minipage}
\includegraphics[width=0.45\textwidth,valign=c]{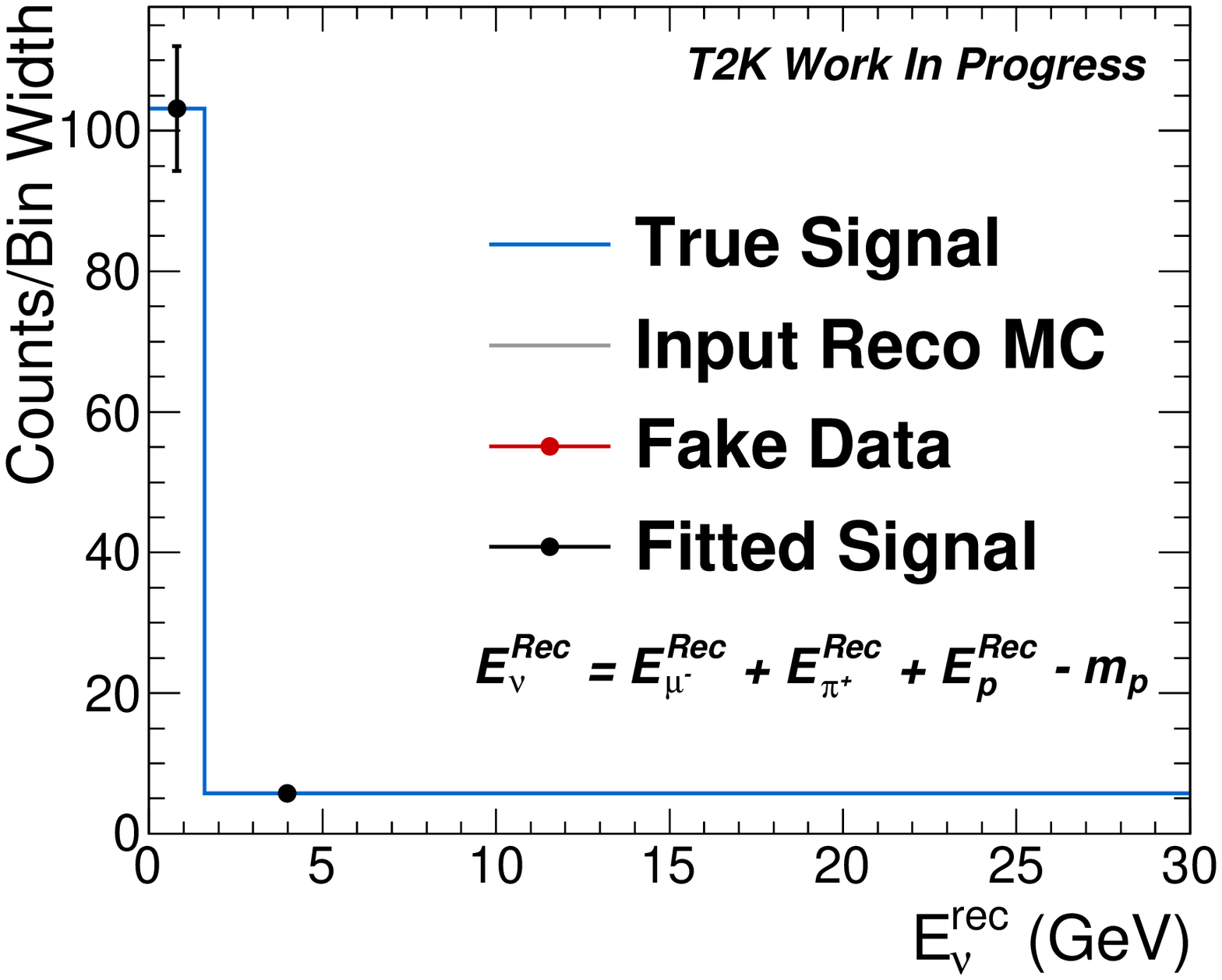}
\caption{Energy dependant fit result where only the statistical term term is considered.}
\label{fig:enuxs}
\end{center}
\vspace{-0.8cm}
\end{figure}

Outlined in~\cite{arXiv:1512.09042}, is a method in which an energy dependent measurement on hydrogen can be achieved. We, for the first time, present such an energy dependent result where only statistical errors are considered. This reduces Eqn.~\ref{eqn:Nexp} to $N^\mathrm{exp}_j = c_iN^\mathrm{MC~signal}_it^\mathrm{det.}_{ij}$. One should note that, there are now two $c_i$ terms to match the number of energy bins. The outcome of the fit can be seen in Fig.~\ref{fig:enuxs}, on the left, the \dptt~fit for each energy bin shows good agreement between the true and fitted result.   The reconstructed energy dependence of the \signal-hydrogen signal is presented in the right of Fig.~\ref{fig:enuxs}. The uncertainty on the $1.6 < E^\mathrm{Rec}_nu (\mathrm{GeV}) < 30$ energy bin is hidden by the bin width normalisation. 
\vspace{-0.2cm}
\section{Conclusion and future work}
The statistical approach defined here, for both an absolute and energy dependent cross section highlight the potential of the \dptt~variable to isolate interactions in hydrogen. Work towards understanding where biases arise when model systematics are included in the fit is ongoing and further evaluation of the dominant systematic, arising from the detector need to be considered. It is hoped that such uncertainties can be reduced by optimising the selection.

\end{document}